\documentclass{PoS}

\title{Using constraint programing to resolve the multi-source / multi-site data movement paradigm on the Grid}

\ShortTitle{Using CP to resolve the multi-source / multi-site data movement paradigm on the Grid}

\author{\speaker{Michal ZEROLA}%
\thanks{The investigations have been partially supported by the IRP
  AVOZ10480505, by the Grant Agency of the Czech Republic under
  Contract No. 202/07/0079, by the grant LC07048 of the Ministry of
  Education of the Czech Republic and by the U.S. Department Of
  Energy.}\\
Nuclear Physics Inst., Academy of Sciences, Prague\\
E-mail:\email{zerola@matfyz.cz}}

\author{J\'{e}r\^{o}me Lauret\\
        Brookhaven National Laboratory\\
        E-mail: \email{jlauret@bnl.gov}}

\author{Roman Bart\'{a}k\\
        Faculty of Mathematics and Physics, Charles University, Prague\\
        E-mail: \email{bartak@kti.mff.cuni.cz}}

\author{Michal \v{S}umbera\\
        Nuclear Physics Inst., Academy of Sciences, Prague\\
        E-mail: \email{michal.sumbera@ujf.cas.cz}}

\abstract{In order to achieve both fast and coordinated data transfer
  to collaborative sites as well as to create a distribution of data
  over multiple sites, efficient data movement is one of the most
  essential aspects in distributed environment. With such capabilities
  at hand, truly distributed task scheduling with minimal latencies
  would be reachable by internationally distributed collaborations
  (such as ones in HENP) seeking for scavenging or maximizing on
  geographically spread computational resources. But it is often not
  all clear (a) how to move data when available from multiple sources
  or (b) how to move data to multiple compute resources to achieve an
  optimal usage of available resources. Constraint programming (CP) is
  a technique from artificial intelligence and operations research
  allowing to find solutions in a multi-dimensional space of
  variables. We present a method of creating a CP model consisting of
  sites, links and their attributes such as bandwidth for grid network
  data transfer also considering user tasks as part of the objective
  function for an optimal solution. We will explore and explain
  trade-off between schedule generation time and divergence from the
  optimal solution and show how to improve and render viable the
  solution's finding time by using search tree time limit,
  approximations, restrictions such as symmetry breaking or grouping
  similar tasks together, or generating sequence of optimal schedules
  by splitting the input problem. Results of data transfer simulation
  for each case will also include a well known Peer-2-Peer model, and
  time taken to generate a schedule as well as time needed for a
  schedule execution will be compared to a CP optimal solution. We
  will additionally present a possible implementation aimed to bring a
  distributed datasets (multiple sources) to a given site in a minimal
  time.}

\FullConference{XII Advanced Computing and Analysis Techniques in Physics Research\\
                 November 3-7, 2008\\
                 Erice, Italy}

\usepackage[utf8]{inputenc}
\usepackage[T1]{fontenc}

\usepackage{amsmath}

\usepackage{graphicx}
\graphicspath{{images/}}

\begin{document}

\section{Introduction}
\subsection{Problem area}
Computationally challenging experiments such as the one from the High
Energy and Nuclear Physics community (HENP) have developed a
distributed computing approach (a.k.a. Grid computing model) to face
the massive needs of their Peta-scale experiments. The era of data
intensive computing has surely opened a vast arena for computer
scientists to resolve practical and exciting problems.  One of such
HENP experiments is the STAR\footnote{http://www.star.bnl.gov}
(Solenoidal Tracker at Relativistic Heavy Ion Collider) experiment
located at the Brookhaven National Laboratory (USA).

In addition to a typical Peta-scale challenge and large computational
needs, this experiment, as a running experiment acquires a new set of
valuable real data every year, introducing other dimension of safe
data transfer to the problem. From the yearly data sets, the
experiment may produce many physics ready derived data sets which
differ in accuracy as the problem is better understood as time
passes. Thus, demands for a large-scaled storage management and
efficient scheme to distribute data grows as a function of time,
while on the other hand, end-users may need to access data sets from
previous years and consequently at any point in time. Coordination is
needed to avoid random access destroying efficiency.

The user's task is typically embarrassingly parallel; that is, a
single program can run $N$ times on fraction of the whole data set
split into $N$ sub-parts without any impact on science reliability,
accuracy, or reproducibility. For a computer scientist, the issue then
becomes how to split the embarrassingly parallel task into N jobs in
the most efficient manner while knowing the data set is spread over
the world and/or how to spread 'a' dataset and best place the data for
maximal efficiency and fastest processing of the task.

The purpose of this work is to design and develop an automated system
that would efficiently use all available computational and storage
resources. It will relieve end users of making decisions among
possible ways of their task execution (which includes locating and
transferring data to desired sites that appear optimal to user) while
preserving fairness. Users' knowledge of the whole system and data
transfer tools will be reduced just to the communication with the
future planner that will guarantee its decision to spread the task and
data sets over chosen sites was, under current circumstances, the most
efficient and optimal.

\subsection{Milestones}
Rather than trying to solve the problem directly from a task
scheduling perspective within a grid environment, we split the problem
into several stages. By isolating data transfer/placement and
computational challenges from each other we get an opportunity to
study the behavior of both sets of constraints separately.

Individual tasks depend on a datasets which size has to be considered
as well, since the time required for its staging and transfers is also
significant.  Therefore, the {\bf first milestone} is to design and
develop the data transfer planner/scheduler. For a given dataset
needed at some site, its aim is to create a plan with an objective to
prepare files from the dataset at a given site within the shortest
time. The next requirement is to define and achieve fair share
transfers within a multiuser environment. This means that if one user
asked for a huge amount of data at some site, then another user who
asked just for one file shouldn't wait until the first user's plan is
finished.

The {\bf next milestone} generalizes data transfer planning between
sites. The goal for this stage is not to transfer files to one
particular site, but do the transfer to several destinations. More
precisely, the planner's goal is to achieve presence of each file
(from user's input task) at one out of all possible destinations,
while still having the objective in mind, to minimize the finish time
of the last file transfer the user waits for.

The second milestone is highly corellated with the {\bf final
  milestone} - scheduling the data transfers together with particular
tasks (jobs) on a grid. The subtask is not finished after a file is
transfered at some destination site, but when the user's job executed
at the same site (and dependent on this file) is finished. Thus, the
planner still has the freedom of choosing a destination site for each
file, but it has to consider that each site has a specific
characteristic of its computational performance. These attributes
include, for example, the number of available CPUs at current site or
the actual load, so it can be more effective to transfer some files
over the slower link to the computationally high performance site (or
vice versa). The final objective is to minimize the finish time of the
last user's job. In this article we focus on the first milestone.

\section{Problem formalization}
In the following part we will present a formal description of the
problem and an approach based on Constraint Programming technique,
used in artificial intelligence and operations research, where we
search for assignment of given variables from their domains, in such a
way that all constraints are satisfied and value of an objective
function is optimal \cite{Marriott98:programming}.

We will introduce the transfer network consisting of sites holding
information which files are available at the site. For each file we
will search for a path leading to the destination and time slots for
each link on transfer path, when a particular file transfer should
occur.

The network consists of a set of nodes $\mathbf N$ and a set of
directed edges $\mathbf E$. The set $\mathbf{OUT}(n)$ consists of all
edges leaving node $n$, the set $\mathbf{IN}(n)$ of all edges leading
to node $n$.  Input received from a user is a set of file names needed
at a destination site $\mathbf{dest}$. We will refer to this set of
file names as to demands, represented by $\mathbf D$.  For every
demand $d$ we have a set of sources ($\mathbf{orig}(d)$), sites where
the file ($d$) is already available.  We will use one decision
variable for every demand and link of the network (edge in graph). The
$\{0,1\}$ variable $X_{de}$ denotes whether demand $d$ is routed over
edge $e$ of the network. The second variable $start_{de}$ denotes
start time of transfer corresponding to the demand $d$ over edge
$e$. More approaches can be found in \cite{conf/cp/Simonis04}.

\begin{equation}
  \label{eq:obj}
  \min_{X_{de}, start_{de}} \max_{e\in \mathbf E}
  \underbrace{\left(start_{de} +
    \frac{size(d)}{speed(e)}\right)}_{end_{de}}\cdot X_{de}
\end{equation}

\begin{equation}
  \label{eq:pcor}
  \forall d\in \mathbf D: \sum_{e\in \cup \mathbf{OUT}(n|n\in
    \mathbf{orig}(d))}X_{de} = 1, \, \sum_{e\in \cup
    \mathbf{IN}(n|n\in \mathbf{orig}(d))}X_{de} = 0
\end{equation}

\begin{equation}
  \label{eq:pcdest}
  \forall d\in \mathbf D: \sum_{e\in \mathbf{OUT}(dest(d))}X_{de} = 0,
  \; \sum_{e\in \mathbf{IN}(dest(d))}X_{de} = 1
\end{equation}

\begin{equation}
  \label{eq:pcoth}
  \begin{split}
  &\forall d\in \mathbf D, \forall n\notin \{orig(d)\cup
    dest(d)\}:\\ \sum_{e\in \mathbf{OUT}(n)}X_{de} \leq 1,& \;
    \sum_{e\in \mathbf{IN}(n)}X_{de} \leq 1, \;\sum_{e\in
      \mathbf{OUT}(n)}X_{de} = \sum_{e\in \mathbf{IN}(n)}X_{de}
  \end{split}
\end{equation}

\begin{equation}
  \label{eq:ic}
  \forall e\in \mathbf E, \forall d\in \mathbf D: X_{de}=1:
  \bigcap[start_{de}, \underbrace{start_{de} +
      \frac{size(d)}{speed(e)}}_{end_{de}}] = \emptyset
\end{equation}

\begin{equation}
  \label{eq:chc}
  \forall n\in {\mathbf N}, \forall d\in \mathbf D: \sum_{e\in
    \mathbf{IN}(n)}\underbrace{\left(start_{de} +
    \frac{size(d)}{speed(e)}\right)}_{end_{de}} \cdot X_{de} \le
  \sum_{e\in {\mathbf{OUT}}(n)}start_{de} \cdot X_{de}
\end{equation}

\begin{displaymath}
  \begin{array}{c}
  X_{de}\in \{0,1\}\\ start_{de} \in {\cal N^{+}}
  \end{array}
\end{displaymath}

The {\it path constraints} (\ref{eq:pcor}, \ref{eq:pcdest},
\ref{eq:pcoth}) state that there is a single path for each demand
(path starting right in one of origin sites, leading to the
destination). Equation (\ref{eq:ic}) ensures there is only one active
file transfer over every edge in time. The last equation states that a
transfer of the file at any site can start only if the file is already
available at the site (Eq. \ref{eq:chc})(i.e., a transfer of the file
to this site has finished). The objective (Eq. \ref{eq:obj}) is to
minimize the latest finish time of transfer over the whole files.


\subsection{Constraint model}
For implementation of the solver we use
Choco \footnote{http://choco.sourceforge.net}, a Java based library
for constraint satisfaction problems (CSP), constraint programming
(CP). Among $70$ available constraints Choco provides also a set for
scheduling and resource allocation, we require most. Closer
illustration of several Choco uses can be found in
\cite{conf/gttse/BenavidesSMC06}, \cite{conf/aict/LazovikAG06}, and
\cite{conf/edoc/WhiteSCWLWF07}. In addition, Java based platform
allows us an easier scheduler integration with currently used tools in
the STAR environment.

Constraints introduced in the previous section were used directly via
appropriate Choco structures, except the equation \ref{eq:ic}, that
ensures at most single file transfer in any time on any link. For
this, we used the {\bf cumulative} scheduling constraint and notation
of tasks and resources. Tasks are represented by their duration, by
ranges for starting and ending times, and by resource consumption
respectively. They are allocated to the resource(s) in such a way that
in any time resource capacity can not be exceeded.

\begin{figure}[h]
  \begin{center}
    \includegraphics[width=4.8in]{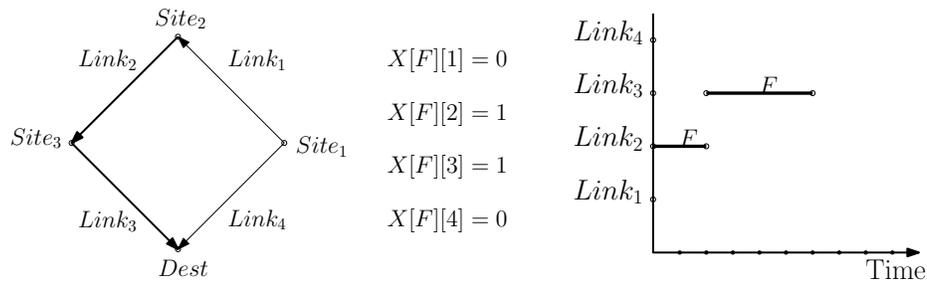}
    \caption{Example of a schedule solution with file $F$ and its
      origin at $Site_1$ and $Site_2$.}
    \label{fig:model_sol_example}
  \end{center}
\end{figure}

In our case, each link acts as a separate resource with capacity $1$
(unary resource) and each file demand creates a single task on every
resource, which duration depends on the current link speed (resource
characteristic) and consumption of the resource corresponds to the
value of variable $X$, i.e. no consumption if the transfer path for
demand does not include current link (resource), or consumption 1
otherwise.  In the Figure (\ref{fig:model_sol_example}) is
shown one possible schedule for transferring one file ($F$) with an
origin at $Site_1$ and $Site_2$ to a destination $Dest$. Values of the
$X$ variables define the path, while the resource profile for each
link is on the right side.

The search strategy, following Choco notation, is split into two {\it
  goals}. First one is to find assignment for $X$ variables,
i.e. paths for each transfer, while the second is to allocate time
slot, assign $start$ variables, for each transfer at chosen links. For
both goals the default {\it `minimum domain`} variable selection and
{\it `increasing value`} value selection heuristic were used.

\section{Direct connections}
In order to closely analyze the problem, its scale, and behavior of
used techniques, we started with several restrictions that simplify
the case. We started to explore the network, where only {\bf direct
  connections} for data movement are allowed. In other words, file can
not be transferred from its origin to the destination by a path longer
than one.

One can think that such a restriction shrinks the search space
enormously, but closer look reveals that the number of possible
combinations is still large:

Let's suppose that we have a network of $5$ sites, all connected to
the destination and $100$ files available at each site
($|orig(f)|=5$). The number of decision variables $X$ is therefore
$500$ ($=|D| * |E|$). Even if an upper bound for all possible
combinations ($2^{500}$) is reduced by a propagation to $5^{100}$
(solver has a freedom of 5 choices of an origin for each file),
brute-force methods can run 'forever'.

With an intend to stay close to a reality, we fixed the number of
sites to $5$, which approximately represents the number of sites
currently available in the STAR experiment. For each link we
introduced a {\it slowdown factor} that influences the transfer time
needed to move the data over this link. Slowdown factor $1$ means that
file of size $1$ unit can be transferred in $1$ unit of time, but with
a slowdown factor $4$ only in $4$ units of time, etc.

Considering the second part of the input, the file demands, we studied
the following cases: a) every file is available only at one particular
site [{\bf distinct}]; b) file is available at sites given by a
probability function, that represents the reality [{\bf weighted}]; c)
file is available at all sites [{\bf shared}].  For all cases we fixed
the file size to a $1$ unit, i.e. all files have the same size.

\subsection{Shared links}
So far we have assumed that all links incoming or outgoing from any
site have their own bandwidth (slowdown factor) that is not affected
by other ones. Nevertheless, in reality this is not always feasible,
since several links leading to a site usually share the same router
and/or physical fiber which bandwidth (capacity) is less than the sum
of their own values. Hence, simultaneously one can't use all links at
their maximum bandwidths.  We express this constraint by adding an
additional resource per each group of shared links. Capacity of the
resource will be the bandwidth of a shared link or a router, while
tasks correspond to the scheduled transfers using any link belonging
to this group with consumptions equal to its slowdown factor.

\subsection{Reducing a search time}
We studied also several techniques for reducing the time spent during
a search.

\subsubsection{Symmetry breaking}
One of the common techniques for reducing the search tree is detecting
and breaking variable symmetries. This is usually done by adding
variable symmetry breaking constraints that can be expressed easily
and propagated efficiently using lexicographic ordering.  One idea
that can be applied in the studied case (direct connections and fixed
file size) is following: if two files have the same origin sets, links
selected for the first one and for the second one respectively must be
ordered. The reason behind is that both files must be transferred to
the destination and their size is equal, it is not necessary to check
also 'swapped' case, since the transfer time can not be shorter.

\subsubsection{Decomposition and search limits}
Another approach is based on the idea, where instead of searching for
a global optimal solution that can be very computing time consuming,
we try to find an optimal solution for smaller parts of the input,
where sum of the time spent will be just a portion of time needed
otherwise. This principle is even more suitable for our needs, since
network link speeds vary in time, some sites can be down after the
schedule is produced, generally, transferring all data files takes
significant amount of time and during this time a lot of factors can
be different to the ones the scheduler considered at the
beginning. Thus the computed optimal schedule for the full input
doesn't have to be valid anymore.

\begin{figure}[h]
  \begin{center}
    \includegraphics[width=4.5in]{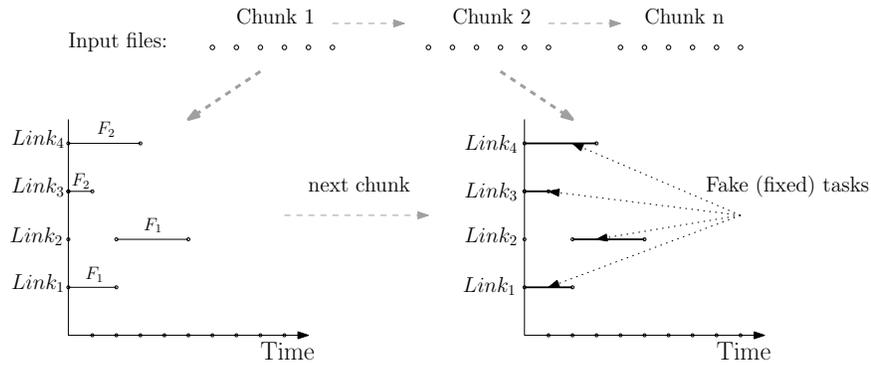}
    \caption{Allocating fake tasks according to the previous schedule.}
    \label{fig:chunk_fake_tasks}
  \end{center}
\end{figure}

One of the approximations is splitting the input files into chunks and
producing an optimal schedule for each chunk separately, while
propagating the results from the previous ones. More precisely, result
of the scheduler for a given chunk of files is information of computed
starting/ending times for each file at particular links. In other
words, current solver receives times for each link, by which the link
will be busy, thus further scheduling for current chunk cannot place
file transfer in these time-slots. We achieve this by allocating a fake
task, with fixed starting and ending times, that were propagated from
previous schedules (Figure: \ref{fig:chunk_fake_tasks}).

Also limits can be imposed on the search algorithm to avoid spending
too much time in the exploration. One of them is fixing the time limit
on a search tree. When the execution time is equal to the time limit,
the search stops whether an optimal solution is found or not. One of
the algorithms we studied was based on this, with a time-limit linearly
dependent on the number of files in a request.

\section{Directed (simple) paths}
Considering the model, no changes are necessary to perform in order to
allow solver search for transfer paths longer than one. However, since
data set transit takes some storage space, one must be sure that
during file transfer from site A to C, using site B, there is enough
space at intermediate site B.

\subsection{Storage capacity}
In order to respect storage restrictions we introduce the next
attribute for each site, the available (free) space, or the storage
capacity. All the time during the execution of a schedule, the storage
capacity constraint for each site must be respected.

\begin{figure}[h]
  \begin{center}
    \includegraphics[width=4.5in]{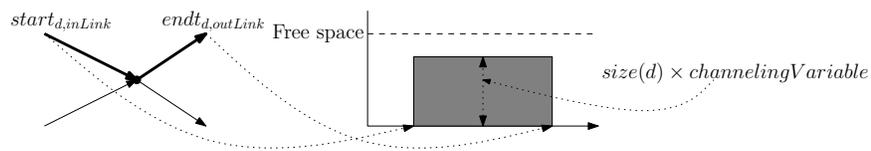}
    \caption{Storage resource and representation of the consumption of
      a free space at site during transferring a file through it.}
    \label{fig:storage_space_resource}
  \end{center}
\end{figure}

For each site we consider all possible ways (pairs of $inLink$ and
$outLink$ how a file can be transferred trough it. Whether or not a
pair is really used for the demand $d$ is expressed by
$channelingVariable$, using which we define also consumption of the
task (Figure: \ref{fig:storage_space_resource}).

If the pair is not used, the consumption is set to zero and storage
resource is invariable to this task, otherwise the consumption is set
to the file size.

\section{Comparative studies}
In this section we present the performance comparison of several
methods of the CSP solver introduced in previous sections as well as
of the Peer-2-Peer simulator. We will also show an effect of one
constraint (storage based) for a simple paths case and an example of
the optimal schedule produced by the solver.

\subsection{Peer-2-Peer simulator}
To provide a base comparison with the results of our CSP based solver
we chose to implement a Peer-2-Peer (P2P) model as well. This model is
well known and successfully used in similar fields like file sharing,
telecommunication, or media streaming. We implemented a P2P simulator
by creating the following work-flow: {\bf a)} put an observer for each
link leading to the destination; {\bf b)} if an observer detects the
link is free, it picks up the file at his site (link starting node),
initiate the transfer, and waits until the transfer is done. We
introduced a heuristic for picking up a file as typically done for
P2P. Link observer picks up a file with a smallest cardinality in the
sense of its $|origin|$, i.e. the file that is available at the
smallest number of sites and if there are more files available with
the same cardinality, it randomly picks any of them. After each
transfer, the file record is removed from the list of possibilities
over all sites. This process is typically resolved using distributed
hash table (DHT) \cite{Naor03}, however in our simulator only simple
structures were used. Finally an algorithm finishes when all files
reach the destination, thus no observer has any more work to do.

\subsection{Results}
\begin{figure}[h]
  \begin{center}
    \includegraphics[width=2.9in]{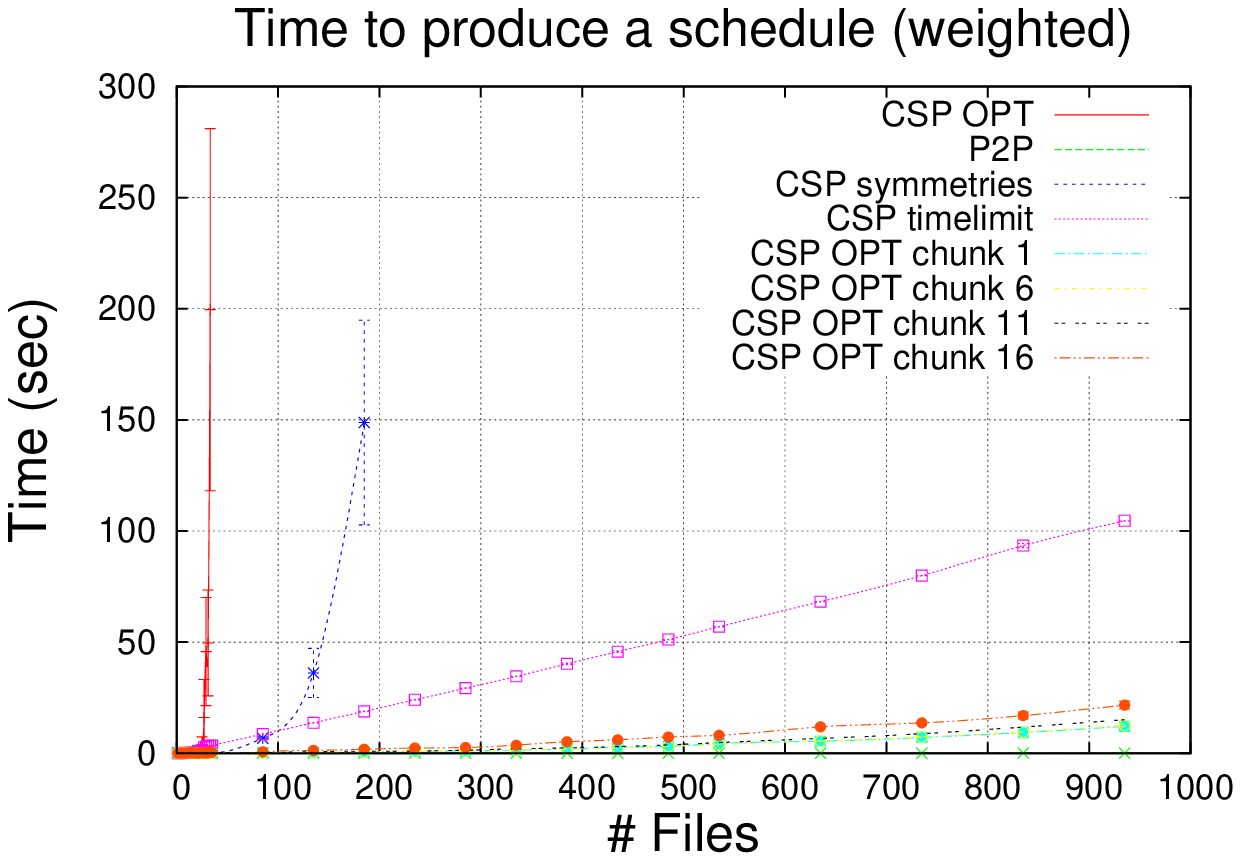}
    \includegraphics[width=2.9in]{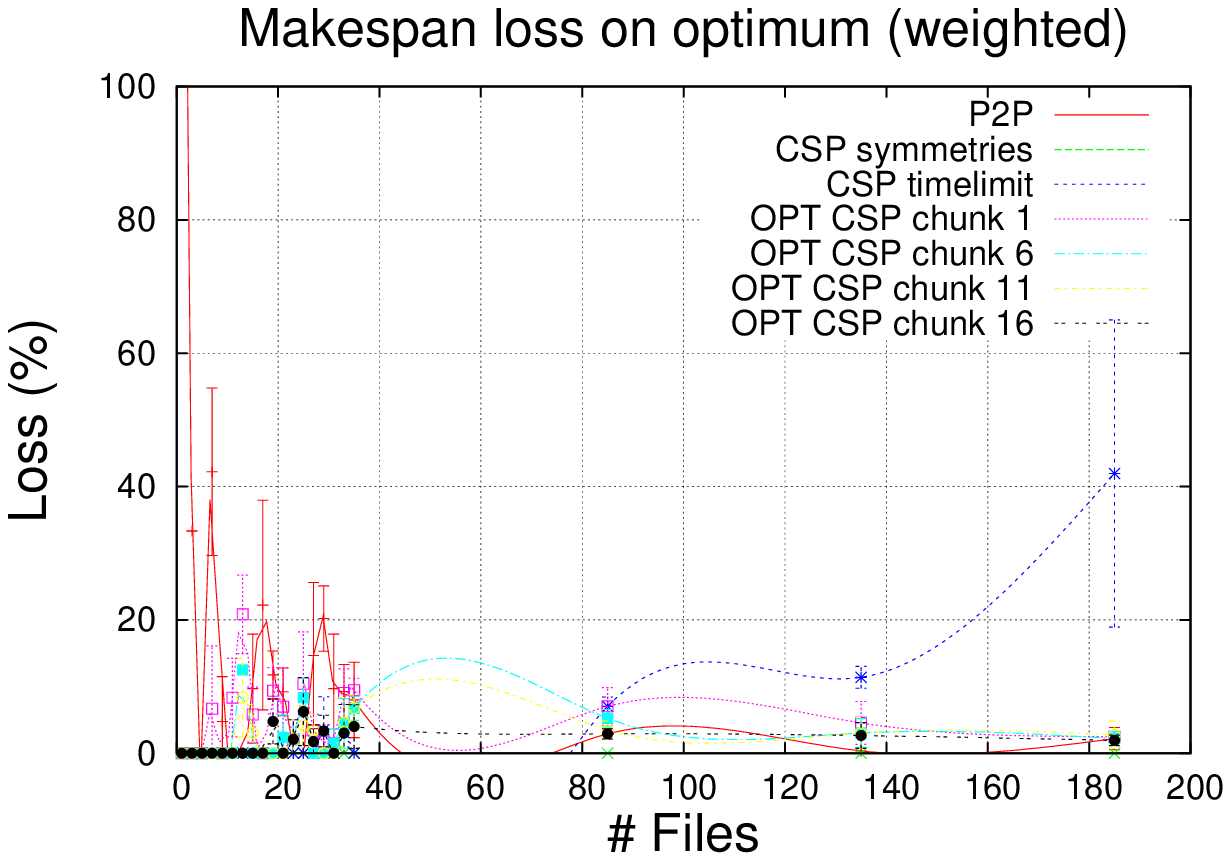}
    \caption{Approximation of the runtime (left) and makespan loss on
      optimal schedule (right) for weighted case.}
    \label{fig:weighted_runtime_makespan}
  \end{center}
\end{figure}

In Figure \ref{fig:weighted_runtime_makespan}, we show a comparison of
times needed to produce the schedules and divergence of the results
(makespan) to the optimal solution between several algorithms. We
present results only for {\bf weighted} case with direct connections
and will only describe the qualitative features for the other
cases. Weights (probabilities) that were used for sites considering
file's origins were $1.0, 0.6, 0.01,$ and $0.01$.

The {\it X} axes denote the number of files in a request while {\it Y}
is the time (in units) needed to generate the schedule and percentage
loss on optimal solution. We can see that time to find an optimal
schedule without any additions grows exponentially and is usable only
for a limited number of files, $50$ in the weighted case and $20$ in
the shared case. This difference is induced by a higher number of
possible configurations as long as any site can be selected as an
origin. By introducing symmetry breaking, the solving time is
improved, but still not usable for more than $200$ files. Using the
time-limit on the other hand we are moving apart from an optimal
solution with increasing files in request, which is even more visible
in shared case. Thus setting the time-limit as a linear function to the
number of files, while using a default search strategy based on
minimal domains, is not sufficient.

In contrast, splitting the input into chunks is giving the best
performance results both in the running time and also in the quality
of the makespan. Even scheduling by chunk of size $1$, i.e. file by
file, doesn't produce worse result than using larger chunks due to
previous conditions propagation. We note as well the efficacious
performance of a simple P2P algorithm, but it is worth to mention that
this model is usable only in a direct connection case, while our
intent is to study more complex networks with much more restrictions.

\begin{figure}[h]
  \begin{center}
    \includegraphics[width=3.8in]{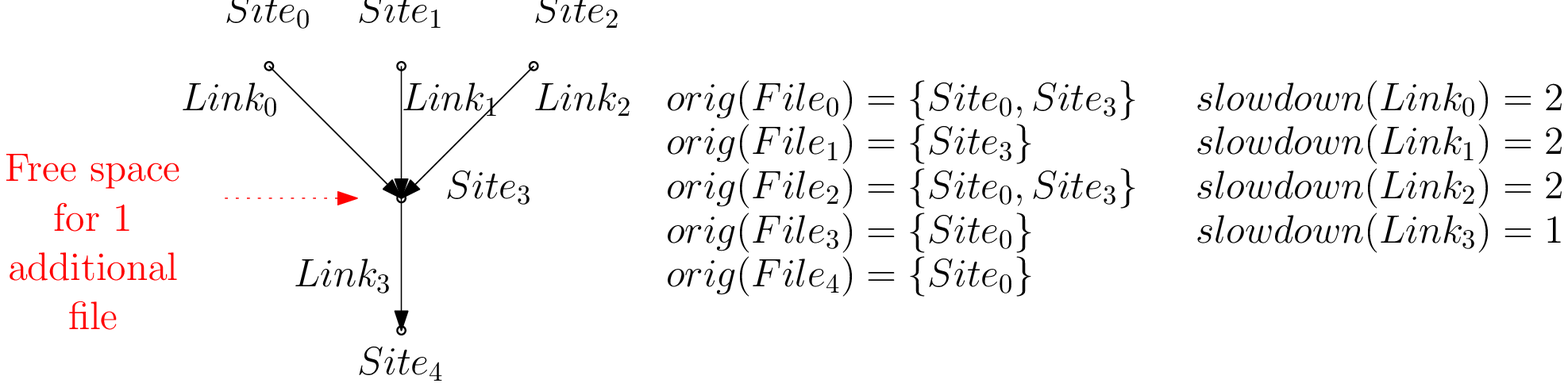}
    \includegraphics[width=2.8in]{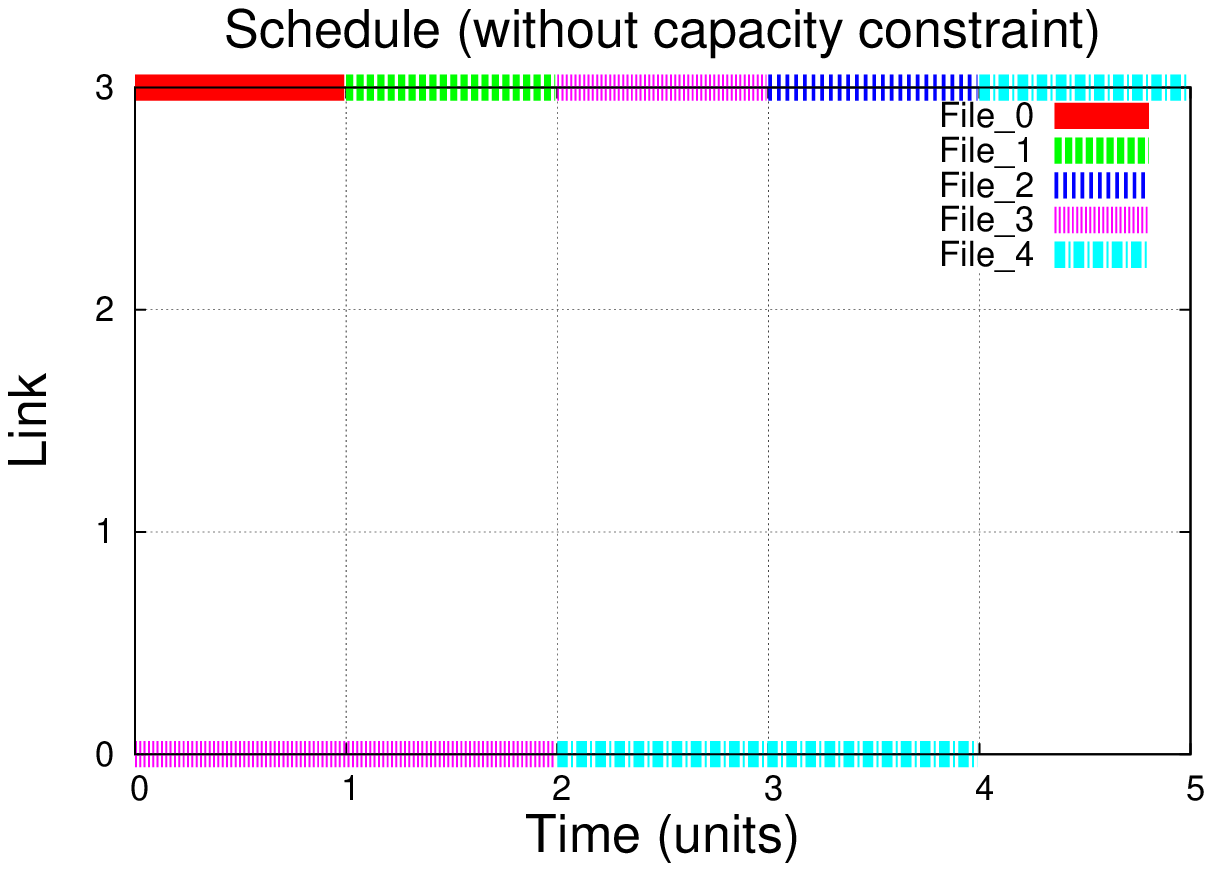}
    \includegraphics[width=2.8in]{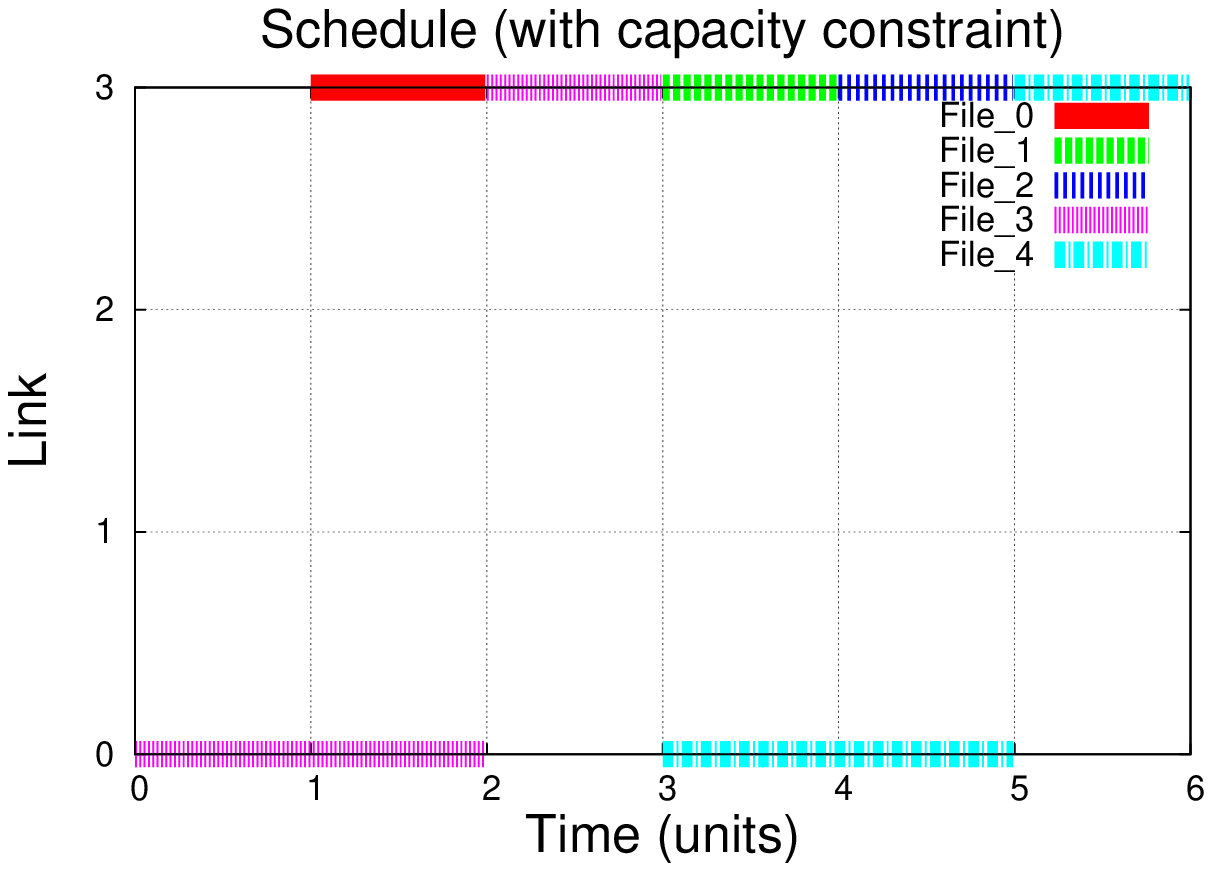}
    \caption{Gantt chart of a schedule without storage constraints
      (left) and a schedule with storage constraints (right) generated
      on the funnel network with limited storage capacity (up).}
    \label{fig:storage_constraints}
  \end{center}
\end{figure}

To see the real effect of the storage constraint, in Gantt charts (Figure
\ref{fig:storage_constraints}) are shown two schedules (without and
with enabled constraint) for the same dataset, considering the funnel
network displayed in the upper part of the figure with a limited
available space at $Site_3$ only for one file size unit. This extreme
example permits only a single transfer via site $Site_3$, that fills
available space until a file is fully transfered to the destination
$Site_4$. After that, the space at $Site_3$ is again released and
another file can go trough.

\section{Conclusion}
We presented an approach using a Constraint Programming model to
tackle the efficient data transfers/placements and job allocations
problem within a distributed environment. Usage of constraints and
declarative type of programming offers straightforward way of
representing many real life restrictions which is also less vulnarable
to dragging bugs in an expanding code. On the other hand, since a
search space is usually extensive, methods like symmetry breaking or
approximations and understanding the scale of the problem are
fundamental. We showed that using the scheduling of data transfers by
sequence of smaller chunks gives results close to the optimal solution
and provides very acceptable running time performance. We have
implemented also several constraints for dealing with shared network
links or limited storage capacities at sites and actual results
indicate that it is worth to continue research with this technique.

\bibliography{zerola_acat_2008}
\bibliographystyle{plain}

\end{document}